	\documentclass[twocolumn,showpacs,preprintnumbers,amsmath,amssymb]{revtex4}

\usepackage{bm}
\usepackage{graphicx}
\usepackage{latexsym}
\usepackage{amssymb}

\renewcommand{\vec}[1]{\mbox{\boldmath$#1$}}
\newcommand{\be}{\begin{equation}}
\newcommand{\ee}{\end{equation}}
\newcommand{\bea}{\begin{eqnarray}}
\newcommand{\eea}{\end{eqnarray}}

\renewcommand{\epsilon}{\varepsilon}
\renewcommand{\phi}{\varphi}

\renewcommand{\div}{\mathrm{\div}}

\renewcommand{\d}{\mathrm{\,d}}
\renewcommand{\v}{\vec}

\newcommand{\R}{\mathbb{R}}

\newcommand{\Int}[1]{\int\limits_{#1}\hspace{-.2cm}}

\newcommand{\ket}[1]{|#1 \rangle}
\newcommand{\braket}[1]{\langle #1 \rangle}

\renewcommand{\em}{\textbf}
\renewcommand{\theta}{\vartheta}

\begin{document}
\bibliographystyle{apsrev}

\title{Density functional theory investigation\\ of antiproton-helium collisions}

	\author{Nils Henkel}
	\email{henkel@pks.mpg.de}
	\affiliation{Max-Planck-Institut f\"ur Physik komplexer Systeme, D-01187 Dresden, Germany}

	\author{Matthias Keim}
	\affiliation{Institut f\"ur Theoretische Physik, Goethe-Universit\"at, D-60438 Frankfurt, Germany}
 
	\author{Hans J\"urgen L\"udde}
	\affiliation{Institut f\"ur Theoretische Physik, Goethe-Universit\"at, D-60438 Frankfurt, Germany}

	\author{Tom Kirchner}
	\affiliation{Department of Physics and Astronomy, York University, Toronto, Ontario, Canada M3J 1P3}

\date{\today}
\date{September 4, 2009}

\begin{abstract}
We revisit recent developments in the theoretical foundations of time-dependent density functional theory (TDDFT). 
TDDFT is then applied to the calculation of total cross sections for ionization processes in the $\overline{p}$-He collision system. The Kohn-Sham potential is approximated as the sum of the Hartree-exchange potential and a correlation potential that was proposed in the context of laser-induced ionization. Furthermore, some approaches to the problem of calculating the ionization probabilities from the density are discussed. Small projectile energies (\,$\le$5\,keV) are considered as well as those in the range from 5 to 1000\,keV. Results are compared with former calculations and with experimental data. We find that the correlation potential yields no obvious improvement of the results over the exchange-only approximation where the correlation potential is neglected. Furthermore, we find the problem of calculating the desired observables crucial, introducing errors of at least the same order of magnitude as the correlation potential. For the case of small energies we find that trajectory effects play an important role: the ionization cross sections are enlarged significantly if curved instead of straight-line trajectories are used for the projectile motion.
\end{abstract}


\maketitle
\section{\label{sec:level1}Introduction}
There are several good reasons for studying atomic collisions. Probably the most important one on the fundamental side is their suitability to shed light on the quantum dynamical many-particle problem. Accordingly, they often serve as
test-beds for many-particle methods. This is especially true for the antiproton-helium system under consideration in the present paper, which is arguably the simplest collision system with more than one electron (Complications associated with electron transfer processes or exchange effects are absent. This is in contrast to bare ion and electron impact.)
This makes it a problem in which on the one hand the interaction between the electrons has to be taken into account, but which, on the other hand, is still simple enough to enable a numerical solution of the time-dependent Schr\"odinger equation with present-day computers. These solutions can then serve as benchmarks for other methods, e.g., time-dependent density functional theory (TDDFT).

In the present paper we explore the TDDFT approach to the problem at hand and are especially concerned with correlation effects. There are two ways in which they are relevant in TDDFT: First, an approximation for the correlation potential is necessary to calculate the time evolution of the density (called dynamical correlation in the following). Second, the observables of interest must be expressed as density functionals (we will call this functional correlation in the following). We investigate a recent model suggested in the context of laser-induced ionization \cite{LK} for the treatment of the dynamical correlation, and one \cite{WB} for the functional correlation, and we give an estimate for the error induced by functional correlation. Furthermore, we will apply these approximations to collisions at lower energies than are usually considered.

In Sec. \ref{theory}, after giving a short introduction to the fundamentals of TDDFT, we introduce the time-dependent exchange-only (hereafter: x-only) approximation investigated in Ref. \cite{Keim} and the above-mentioned approximations for correlation effects. 
A few remarks on the computational methods used are also provided. In Sec. \ref{Results} we present our results and compare them with experimental data and several other calculations. Conclusions are offered in Sec. IV. Atomic units are used unless indicated otherwise.

\section{Theory and Computations}\label{theory}

\subsection{Time-dependent density functional theory (TDDFT)}
TDDFT is a method for the investigation of systems of $N$ indistinguishable, interacting particles, described by a time-dependent Schr\"odinger equation (TDSE) 
\begin{equation}
  i\partial_t\ket{\Psi(t)} = \hat H(t)\ket{\Psi(t)}\ \forall\ t\in[0,T],\enskip \ket{\Psi(0)}=\ket{\Psi_0}
\end{equation}
with the Hamiltonian
\begin{equation}
 \hat H(t) = \hat T+\hat W+\hat V(t),
\end{equation}
where $\hat T$ is the kinetic energy operator, $\hat W$ is the interaction potential and $\hat V$ an external potential that depends on time. We denote by $\v r_l$ the position, by $\sigma_l$ the spin of the $l$-th particle and by $\v x_l$ the combination of both, and we set $\v x=(\v x_1\dots \v x_N)$.\\
The first Runge-Gross theorem \cite{RG} states, that if the initial state $\ket{\Psi_0}$ and the interaction potential $\hat W$ are fixed, the time-dependent (one-particle) density function $n:\R^4\rightarrow\R^+$, defined by
\begin{equation}
 n(\v r_1,t) = N\sum_{\boldsymbol\sigma}\int\d^3r_2\dots\int \d^3r_N|\Psi(\v x)|^2
\end{equation}
 determines $\hat V$ up to a time-dependent constant.
Hence, all physical information is determined as it is invariant under the transformation $\hat V(t)\rightarrow \hat V(t) + c(t)$. Especially, the state of the system is determined for all times up to a merely time-dependent phase. We denote the state that yields the density $n$ for the interaction $\hat W$ by $\ket{\Psi[\hat W, n]}$.

$n$ can be obtained by solving the so-called Kohn-Sham equations, which are one-particle TDSEs for noninteracting particles in an effective, density-dependent, external potential $\hat V^{\rm{KS}}[n]$:
\begin{equation}
 i\partial_t\ket{\phi_l(t)} = \hat H^{\rm{KS}}[n](t)\ket{\phi_l(t)}, \enskip\ket{\phi_l(0)}=\ket{\phi_l^0}
\end{equation}
with $\hat H[n]^{\rm{KS}}(t) = \hat T+\hat V^{\rm{KS}}[n](t)$ and the one-particle states $\ket{\phi_1}\dots\ket{\phi_N}$.\\
Originally (in \cite{RG}) $\hat V^{\rm{KS}}[n]$ had been derived from the Frenkel-stationarity principle as 
\begin{equation}
 V^{\rm{KS}}[n](\v r_1,t) = V(\v r_1,t) + \frac{\delta \mathcal{A}_0[\hat W, n]}{\delta n(\v r_1,t)} - \frac{\delta \mathcal{A}_0[0, n]}{\delta n(\v r_1,t)}\label{KS_pot_ori}
\end{equation}
where
\begin{equation}
 \mathcal{A}_0[\hat W, n]=\int_0^T\d t\,\braket{\Psi[\hat W, n]|i\partial_t-\hat T-\hat W|\Psi[\hat W, n]}\label{action}
\end{equation}
is the quantum mechanical action. 
Note that it has remained a controversial issue if the potential $V^{\rm{KS}}[n](\v r_1,t)$ at a certain time $t$ can depend on the density   $n(\v r_1,t')$ at a later time $t'>t$. If that is the case, the problem has to be solved by a self-consistency iteration. Recent discussions \cite{SchirmerKritik}, \cite{MaitraKommentar}, \cite {SchirmerReply} were concerned with the question whether such an iteration is indeed necessary and whether its convergence can be guaranteed. In the approximations we use, however, this is not an issue as will be seen in Sec. \ref{corrpot}.

As it turned out \cite{DFT1996}, the definition (\ref{KS_pot_ori}) leads to a violation of causality. This means that the derivation from the Frenkel-stationarity principle must be incorrect. Interestingly, while the violation of causality has been well-known for some time, until recently there was no paper that pinpointed the error in the derivation and addressed the question whether it can be corrected.
This void has now been filled by Vignale who showed that the Frenkel principle is (in its usual form) not applicable in TDDFT \cite{Vignale}.
We would like to rephrase the reason why the original derivation of Ref. \cite{RG} went wrong:
Let $\ket{\tilde\Psi}$ be any state and let $\ket{\Psi}$ be the solution of the TDSE with the Hamiltonian $\hat H$ and a given initial condition. Frenkel's stationarity principle states that $\ket{\tilde\Psi}=\ket{\Psi}$ if and only if
\begin{itemize}
 \item $\ket{\tilde\Psi(0)}=\ket{\Psi(0)}$ and $\ket{\tilde\Psi(T)}=\ket{\Psi(T)}$ and
 \item the action is stationary at $\ket{\tilde\Psi}=\ket{\Psi}$.
\end{itemize}
However, when one translates this principle from states to densities, the validity of the final condition can never be ensured: In TDDFT $\ket\Psi$ is not known, but only $n$. While the Runge-Gross theorem states that $n$ determines $\ket\Psi$ it is not true, that $n(.,T)$ determines $\ket{\Psi(T)}$, instead $n$ would have to be known at all times. Hence, the boundary condition translates to ``$n(.,t)$ is the true density of the system at all times'', making the principle a meaningless tautology.
One is inclined to think that the same argument might hold for the initial condition. This is true in general, however, if it is \textit{additionally} assumed that $\ket{\Psi(0)}$ is the ground state of the initial Hamiltonian, then ground-state DFT guarantees that $n(.,0)$ determines $\ket{\Psi(0)}$. 

This explains why the original result of Runge and Gross violates causality. It also shows a way out: Leave away the final condition. This is exactly what was done by Vignale and which leads to the appearance of additional terms in the Kohn-Sham potential compared to Eq. (\ref{KS_pot_ori}) (see Ref. \cite{Vignale} for details).

\subsection{The antiproton-helium system}
The system under consideration in the present paper consists of an antiproton, passing on a classical path by a para-helium atom (where electrons have antiparallel spins). The impact parameter is $b$, the antiproton's kinetic energy in the laboratory system (the system where the helium atom is at rest at the initial time) is $E$. For sufficiently high energies a straight-line trajectory for the antiproton can be assumed. At the initial time the (spatial) Kohn-Sham state can be written as $\ket\Phi=\ket{\phi}\ket{\phi}$ (using the abbreviation $\ket{\Phi[n]}=\ket{\Psi[0,n]}$); the total state is antisymmetrical due to the spin function. We neglect all spin-dependent interactions, so the spin function can be ignored. Since the Kohn-Sham Hamiltonian contains no interaction term, the two Kohn-Sham equations of the system as well as the corresponding initial conditons are identical and the problem is reduced to one and only one one-particle TDSE
\begin{equation}
 \hat H^{\rm{KS}}[n]\ket{\phi(t)} = i\partial_t\ket{\phi(t)}.\label{KS-eq}
\end{equation}
Unfortunately, no exact formula for the Kohn-Sham potential is known, therefore, approximations are necessary. First, the potential is split up:
\begin{equation}
 \hat V^{\rm{KS}}[n] = \hat V +\hat V_{Hx}[n] + \hat V_c[n],\label{KS_pot}
\end{equation}
where $\hat V$ denotes the external potential by the antiproton und the helium nucleus. The so-called Hartree-exchange potential is defined by
\begin{equation}
 V_{Hx}[n](\v r_1,t) = \frac{\delta \int_0^T\d t\,\braket{\Psi[0, n]|\hat W|\Psi[0, n]}}{\delta n(\v r_1,t)},
\end{equation}
which (in our case) gives
\begin{equation}
 V_{Hx}[n](\v r_1,t) = \frac12\int\d^3r_2\ n(\v r_2)W(\v r_1,\v r_2).\label{Hx-pot}
\end{equation}

 $\hat W$ is the Coulomb interaction between two electrons, i.e., $W(\v r_1,\v r_2)=\tfrac{1}{|\v r_1-\v r_2|}$. The correlation potential $\hat V_c$ is unknown, and approximations are discussed in the subsequent section.

\subsection{The correlation potential}\label{corrpot}
\begin{figure}[htbp]
	\centering
	\includegraphics[width=.35\textwidth]{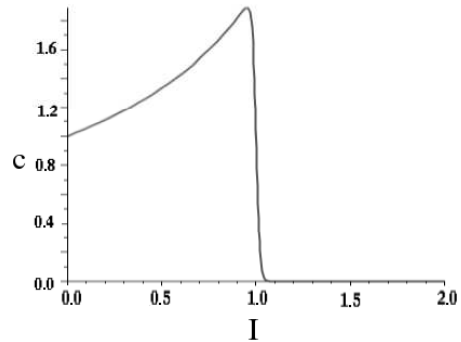}
	\caption{Switch function $c$ [Eq.~(\ref{eq:switch})] as a function of the ionization $I$.}
	\label{switch}
\end{figure}

One possible approximation for the Kohn-Sham potential (\ref{KS_pot}) is the so-called x-only approximation, where the correlation potential is neglected: $\hat V_c=0$. This was investigated in Ref. \cite{Keim} and will serve as one of our references. An explicit model for $\hat V_c$ was proposed in the context of laser-driven ionization by Lein and K\"ummel\cite{LK}:
\begin{equation}
 \hat V_{c}(t)=\left[c\left(\tfrac{2}{2-I(t)}\right)-1\right]\cdot \hat V_{Hx}(t).\label{3.14}
\end{equation}
Here, $c$ is a switch function (see also Fig.\ref{switch}):
 \begin{equation}
 c(x)=\frac{x}{1+e^{50x-100}},\label{eq:switch}
\end{equation}
and $I$ is the ionization (average number of electrons in scattering states), i.e.,
\begin{equation*}
 I = 2 - \lim\limits_{R\rightarrow\infty}\lim_{t\rightarrow\infty}\Int{S(R)}\hspace{-.2cm}\d^3 r_1\ n(\v r_1,t),\label{Ionisation}
\end{equation*}
where $S(R)$ is a sphere around the target with radius $R$. We will call this approximation for the correlation potential the LK (Lein and K\"ummel) approximation and the resulting Kohn-Sham potential the LK KS potential.

The approximation assumes no static correlation (i.e., the initial state is identical to the initial state in the x-only approximation) and mimics a discontinuity in the effective electron-electron potential: when $I$ reaches $1$, $c$ becomes zero very quickly and therefore 
\begin{equation}
\hat V_{Hx} + \hat V_c = 0.
\end{equation}
That means that the electron-electron interaction is switched off as soon as (on average) one electron is ionized. \\
Since $\hat V_{Hx}$ is local in time both the x-only and the LK KS potentials are local in time and an iteration procedure is not necessary.
Instead, the Kohn-Sham equation can be solved by standard time propagation.

\subsection{The observable problem}\label{ObsProblem}
Even if the correlation potential (and, therefore, the whole Kohn-Sham Hamiltonian) were known exactly, the calculation of observables would still be a non-trivial problem. While the Runge-Gross theorem guarantees that the density determines all observables, exact formulas for their calculation are known only in very few cases \cite{Luedde}.
The observables we are mainly interested in in this work are the $q$-fold ionization probabilities $p^{q+}(b)$ (dependent on the impact parameter $b$) and the corresponding total cross sections (TCS) $\sigma^{q+}=\int_0^\infty\d b\,bp^{q+}(b)$. They fall into the category of not exactly known observables (i.e., observables whose functional dependence on the density is not known).

\subsubsection{Binomial approximation}
Assume that the Kohn-Sham state $\ket\Phi = \ket\phi\ket\phi$ were the true state of the system and let $p_s$ be the ionization probability of the particle described by $\ket\phi$. Then the two-particle probabilities follow simply from binomial statistics as
\begin{eqnarray}
p_s^{0+}&=&(1-p_s)^2,\nonumber \\
p_s^{1+}&=&2p_s(1-p_s), \label{binomial} \\
p_s^{2+}&=&p_s^2 \nonumber. 
\end{eqnarray}
As we will see, this a rather rough approximation.

\subsubsection{Simplified adiabatic approximation}
One can define the correlation integral 
\begin{equation}
 \mathfrak{I}_c = \lim\limits_{R\rightarrow\infty}\lim_{t\rightarrow\infty}\Int{S(R)^2}\hspace{-.2cm}\d^6 r\ \Bigl(\rho(\v r_1,\v r_2,t)-\tfrac12n(\v r_1,t)n(\v r_2,t)\Bigr),
\end{equation}
where $\rho$ is the two-particle density. With $\mathfrak{I}_c$ the exact probabilities can be expressed as \cite{WB}
\begin{eqnarray}
  p^{0+} &=& p_s^{0+}+\tfrac12\mathfrak{I}_c,\nonumber \\
 p^{1+} &=& p_s^{1+}-\mathfrak{I}_c,\label{ww-p} \\
 p^{2+} &=& p_s^{2+}+\tfrac12\mathfrak{I}_c.\nonumber
\end{eqnarray}
Since the functional dependence of $\rho$ on $n$ is unknown, $\mathfrak{I}_c$ cannot be calculated exactly. An approximation was introduced by Wilken and Bauer (WB) in Ref. \cite{WB}. They rewrote $\mathfrak{I}_c$ as
\begin{equation}
 \mathfrak{I}_c = \lim\limits_{R\rightarrow\infty}\lim_{t\rightarrow\infty}\Int{S(R)^2}\hspace{-.2cm}\d^6 r\ g_c(\v r_1,\v r_2,t)n(\v r_1,t)n(\v r_2,t) \label{eq:Ic}
\end{equation}
with the correlation function 
\begin{equation}
 g_c(\v r_1,\v r_2,t)=\frac{\rho(\v r_1,\v r_2,t)}{n(\v r_1,t)n(\v r_2,t)}-\frac12 , \label{eq:gc}
\end{equation}
and then replaced $n$ and $\rho$ in $g_c$ by the adiabatic approximations
\begin{align*}
 n^A &= \begin{cases}
		In_1 + (1-I)n_2 & \mbox{for } 0\leq I\leq 1\\
                (2-I)n_1 & \mbox{for } 1\leq I\leq 2
                 \end{cases},\\
 \rho^A &= \begin{cases}
		(1-I)\rho_2 \hspace{1.2cm}& \mbox{for } 0\leq I\leq 1\\
                0 & \mbox{for } 1\leq I\leq 2
                 \end{cases},
\end{align*}
where $n_1$ is the ground-state density of $\rm{He}^{1+}$, $n_2$ that of He and $\rho_2$ ist the two-particle ground-state density of He \footnote{Note that if one would replace $n$ by its adiabatic approximation not only in Eq.~(\ref{eq:gc}), but also in Eq.~(\ref{eq:Ic}) one would obtain $\mathfrak{I}_c=-1/2 I^2$ and $p^{2+}=0$ for $ 0\leq I\leq 1$. Similarly, for $ 1\leq I\leq 2$ one would obtain $\mathfrak{I}_c=-1/2 (2-I)^2$ and $p^{0+}=0$.}.

\begin{figure}[hbp]
	\centering
	\includegraphics[height=.4\textwidth, angle=-90]{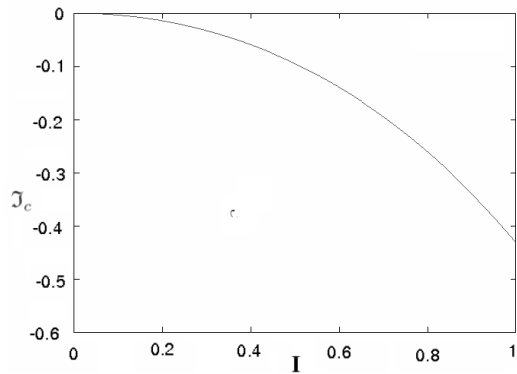}
	\caption{Simplified version of the adiabatically approximated correlation integral [according to Eq. (\ref{sWB})] as a function of the ionization $I$. }
	\label{KorrInt}
\end{figure}

We will use a strongly simplified version of this approximation: The right panel of Fig. 3 in Ref. \cite{WB} shows the dependence of $\mathfrak{I}_c$ on the ionization $I$ of a laser-driven model helium atom for two different sets of laser parameters. In both cases, the dependence is almost identical and a polynomial regression yields (see also Fig. \ref{KorrInt}):
\begin{equation}
 \mathfrak{I}_c(I) = -0.352I^2 - 0.0296I^3 - 0.0502I^4\quad\mbox{for } 0\leq I\leq 1.\label{sWB}
\end{equation}
In our context the case $I>1$ is irrelevant since such large ionizations do not arise. We will call the model that uses the analytical expression (\ref{sWB}) in (\ref{ww-p}) the sWB-approximation.

\subsubsection{Average ionization and bounds for the probabilities}
The observable problem can be circumvented by not trying to calculate the probabilities, but only the average ionization $I$ which can be written as
\begin{equation}
I = p^{1+} + 2p^{2+}.
\end{equation}
From Eq.~(\ref{ww-p}) it follows immediately that the ionization can be calculated exactly from the non-interacting probabilities:
\begin{equation}
 I = p_s^{1+} + 2p_s^{2+}.
\end{equation}
If one wants to benchmark a certain approximation to the Kohn-Sham potential, the average ionization is the observable that should be used because it can be calculated exactly from the density.
Furthermore, $I$ is obviously an upper bound for $p^{1+}$ and for $p^{2+}$ (however, in our system $p^{1+}\gg p^{2+}$, so it is a sensible bound only for $p^{1+}$). There is no known exact lower bound for $p^{1+}$, but one can find arguments that for not too high impact energies $p^{1+}>p_s^{1+}$:\\
\begin{itemize}
 \item The WB-approximation as well as the simplified version yield $\mathfrak{I}_c(I)<0$.
 \item Eq.~(\ref{binomial}) implies that $p_s^{1+}\leq\tfrac12$, but this is not true for $p^{1+}$.
\end{itemize}
Applying these bounds, the probability can be estimated as
\begin{equation}
p^{1+} = \left(p_s^{1+} +p_s^{2+}\right)\pm p_s^{2+}.\label{observable-problem}
\end{equation}

\subsection{Computational details}\label{CompDet}
The initial condition is the x-only Kohn-Sham ground state of para-helium \cite{EngelVosko}. As mentioned above, the time-locality of our approximations of the Kohn-Sham potential allows us to solve the Kohn-Sham equation by incrementally propagating this state. To this end the state and the density are expressed in a basis generated by the basis generator method \cite{BGM}. This basis consists of bound states (atomic orbitals) up to the principal quantum number $n=4$ and of pseudo scattering states constructed from the bound states as described in Ref. \cite{Keim}. We use the same basis as was used in that work and also the same technique and parameters to calculate the average ionization $I$ and the Kohn Sham potential in every time step.

\section{Results}\label{Results}
\subsection{Medium and high energies}

\begin{figure}[hbp]
	\centering
	\includegraphics[height=.4\textwidth, angle=-90]{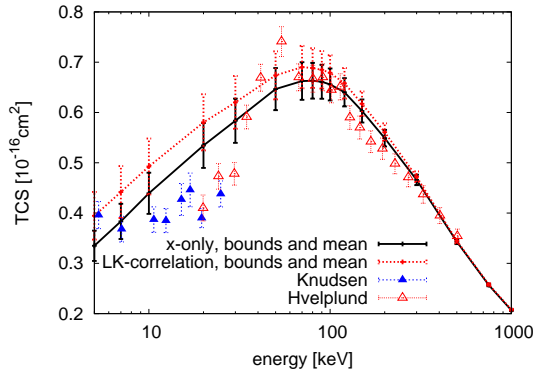}
	\caption{(Color online)  Total cross section for one-fold ionization as a function of projectile energy. Present results within x-only and LK approximations with error bars denoting the bounds according to Eq. (\ref{observable-problem}). Experiments: Knudsen \cite{KnudsenTCS_1}, Hvelplund \cite{Hvelplund}. }
	\label{TCS_1_Schranken}
\end{figure}

\begin{figure}[hbp]
	\includegraphics[height=.4\textwidth, angle=-90]{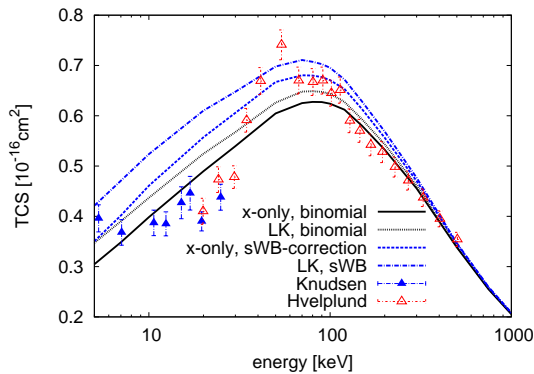}
	\caption{(Color online) Total cross section for one-fold ionization as a function of projectile energy. Present results with various models for the Kohn-Sham potential (x-only, LK) and various approaches for the calculation of probabilities (binomial, sWB). Experiments: Knudsen \cite{KnudsenTCS_1}, Hvelplund \cite{Hvelplund}.}
	\label{TCS_1_Korrekturen}
\end{figure}

First, we compare our results with experiments by Knudsen et al. \cite{KnudsenTCS_1} and Hvelplund et al. \cite{Hvelplund} (Figs. \ref{TCS_1_Schranken} and \ref{TCS_1_Korrekturen}). The figures show $\sigma^{1+}$ as a function of the projectile energy. In Fig. \ref{TCS_1_Schranken}, the observable problem is considered by using the bounds according to Eq. (\ref{observable-problem}) and interpreting $p_s^{1+}+p_s^{2+}$ as the best approximation for $p^{1+}$. In Fig. \ref{TCS_1_Korrekturen} binomial and sWB approximations are used.

The x-only approximation and the LK-approximation coincide for high energies. This is because for high energies there is not enough time for any change in the effective electron-electron-potential to make an important contribution. Furthermore, both models coincide with the experimental data. For small energies we see differences. The correlation enlarges the cross section for a simple reason: In the antiproton-helium system an ionization larger than 1 is almost never reached, and for $I<1$ the correlation potential simply amplifies the Hartree-exchange potential, which is repulsive.
For a closer investigation of the correlation potential the helium atom has to be perturbed more strongly, e.g., by a multiply charged ion. This would lead to higher average ionizations.

The effect of the correlation potential on the cross sections is of the same order of magnitude as the possible deviations induced by the observable problem. Except in the range from 10 to 30\,keV, our results are not in contradiction with the results by Knudsen et al. \cite{KnudsenTCS_1} or Hvelplund et al. \cite{Hvelplund}. However, the uncertainties in our results due to the observable problem are quite large, so a strong agreement cannot be claimed either, which makes a more definite statement impossible.

\begin{figure}[htbp]
	\centering
	\includegraphics[height=.4\textwidth, angle=-90]{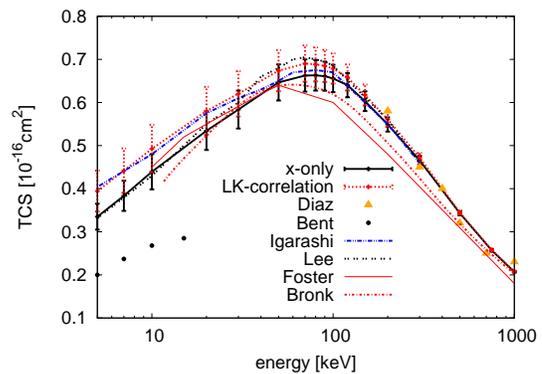}
	\caption{(Color online) Total cross section for one-fold ionization as a function of projectile energy. Present results within x-only and LK approximations with error bars denoting the bounds according to Eq. (\ref{observable-problem}). Two-electron calculations: Diaz \cite{Diaz}, Bent \cite{Bent}, Igarashi \cite{Igarashi}, Lee \cite{Lee}, Foster \cite{Foster}, Bronk \cite{Bronk}.}
	\label{TCS_1_Rechnungen} 
\end{figure}

Supplemental to the comparison with experimental data, we compare our calculated TCS to various theoretical investigations of the system in question in Fig. \ref{TCS_1_Rechnungen}. 
From the large body of published results we have included only those, which were obtained from a solution of the two-electron TDSE, i.e., which include correlation effects. A more comprehensive list of previous calculations can be found in Ref. \cite{KnudsenTCS_1}.
At high energies all calculations, except that by Foster et al. \cite{Foster}, coincide. This agreement can be explained by perturbation theory. For small energies, the calculation by Foster et al. and that by Lee et al. agree with our x-only approximation, while the one by Igarashi et al. agrees with our LK approximation. This leaves the question open, whether the LK approximation improves upon the x-only approximation.

\begin{figure}[tbp]
	\centering
	\includegraphics[height=.4\textwidth, angle=-90]{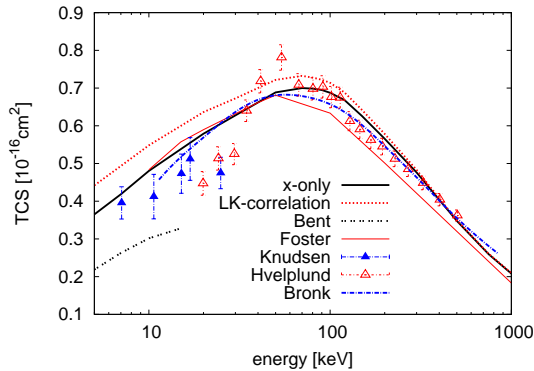}
	\caption{(Color online) Total cross section for average ionization as a function of projectile energy. Present results within x-only and LK approximations. Two-electron calculations: Bent \cite{Bent}, Foster \cite{Foster}, Bronk \cite{Bronk}. Experiments: Knudsen \cite{KnudsenTCS_1} and \cite{KnudsenTCS2}, Hvelplund \cite{Hvelplund}. }
	\label{TCS_Erwartung}
\end{figure}

As mentioned in Sec. \ref{ObsProblem}, the average ionization is actually better suited for comparisons with experimental data and many-particle calculations since it can be expressed as a simple density functional and is only sensitive to the approximation used for the KS potential. Figure~\ref{TCS_Erwartung} shows $\sigma=\sigma^{1+}+2\sigma^{2+}$ as a function of the impact energy. Again, good agreement between our calculations, experiment and the calculation by D\'{\i}az et al. \cite{Diaz} can be seen. For small energies there is good agreement between the x-only approximation and the calculation by Foster et al. (while this agreement can already be seen in the one-fold ionization, it is not conclusive there because of the large uncertainties induced by the observable problem). This is another hint (additional to the fact that $I<1$) that the x-only approximation is better suited for our system than the LK approximation.  The reasons for the deviations between the results of our x-only calculation and that by Foster et al. calculation for medium and high energies are not clear, but the good agreement of our results with the other shown data suggests that our results are solid in that region.

\begin{figure}[tbp]
	\centering
	\includegraphics[height=.43\textwidth, angle=-90]{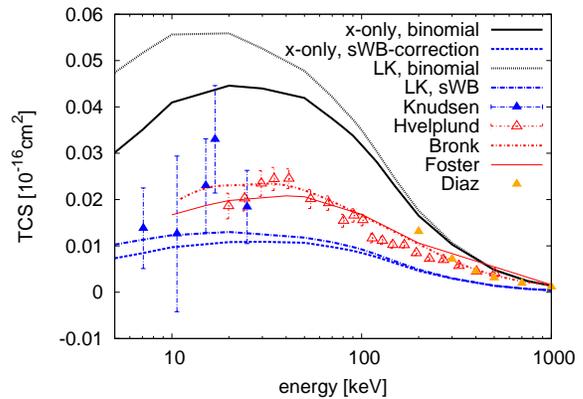}
	\caption{(Color online) Total cross section for two-fold ionization as a function of projectile energy. Present results with various models for the Kohn-Sham potential (x-only, LK) and various approaches for the calculation of probabilities (binomial, sWB). Two-electron calculations: Bronk \cite{Bronk}, Foster \cite{Foster}, Diaz \cite{Diaz}. Experiments: Knudsen \cite{KnudsenTCS2}, Hvelplund \cite{Hvelplund}.}
	\label{TCS_2}
\end{figure}

While no reasonable bounds can be given for the two-fold ionization probabilities, the approximations (\ref{binomial}), and (\ref{sWB}) in (\ref{ww-p}) can be applied, yielding the results displayed in Fig.~\ref{TCS_2}. Once again, x-only and LK approximations agree at high energies, while the TCS obtained from the LK approximation is higher at smaller energies. This confirms what has already been learned from the one-fold ionization. It can be seen that for two-fold ionization the observable problem is even more important than the question of the right approximation for the correlation potential.
These large deviations introduced by the observable problem have to be considered a setback for the TDDFT approach. In order for TDDFT to be a viable approach for the calculation of two-fold ionization it would be necessary to find a much better approximation for the probability as a functional of the density than that which we applied.

\subsection{Low energies}
So far, we investigated impact energies larger than 5\,keV. However, lower energies are in principle accessible experimentally (at CERN) and might prove to be interesting.
Straightforward application of our method leads to a numerical problem concerning the scattering states: While they are constructed in a way that is suited to describe whether an electron is in a scattering state or not, the dynamics of an electron in a scattering state are not described well. Especially, the average electron-target distance is strongly underestimated and the electrons are found to be in close vicinity of the target. When the density obtained from such a state is used to calculate the Kohn-Sham potential, those electrons screen the target potential very strongly, leading to a much too high ionization probability. At sufficiently high energies this numerical effect is of no great concern since the time development is stopped before the effect grows large enough.

For the results presented in this section we circumvent this problem by calculating the Kohn-Sham potential not from the full density, but only from that part obtained from the bound states: Let $\{\ket{\phi_i^b}\}$ be the bound basis states, $\{\ket{\phi_j^f}\}$ the scattering states. The full density is
\begin{equation}
 n(\v r_1,t) = 2\left|\sum_ia_i(t)\phi_i^b(\v r_1) + \sum_jb_j(t)\phi_j^f(\v r_1)\right|^2.
\end{equation}
We define the bound density by
\begin{equation}
 n^b(\v r_1,t) = 2\left|\sum_ia_i(t)\phi_i^b(\v r_1)\right|^2.
\end{equation}
In the same way a free density $n^f$ can be defined (note, however, that in general $n\neq n^f+n^b$). Calculating the potential by using only $n^b$ is equivalent to the assumption that an electron in a scattering state moves away very quickly and has no further influence on the problem.

In the low energy range, the straight-line approximation for the projectile is no longer justified. Instead we use a classical trajectory determined by the initial kinetic energy and the approximate force by the target nucleus and the electrons in the following way:
We assume the target (including the electrons) to be a point charge with the time-dependent effective charge
\begin{equation}
 Q_{eff}(t) = 2 - \Int{S(R(t))}\d^3r_1\, n(\v r_1,0), \label{eq:qeff}
\end{equation}
where $R(t)$ is the distance between the projectile and the target nucleus at time $t$. That means that the nuclear charge of $2$ is reduced by the electrons that were initially closer to the target nucleus than the projectile is at the considered time \footnote{We did not use the full time-dependent density in Eq.~(\ref{eq:qeff}) because of the technical problems mentioned above. However, we also mimicked this dynamical effect by weighing the integral with the time-dependent ionization $I$ and obtained practically the same results.}. It is not taken into account that the target system is no longer inertial when the force between target and projectile is nonzero. This would lead to an additional term in the TDSE \cite{LandauL}. In a classical picture this effect can be described as the nucleus being pulled away from under the electrons which remain in place due to their inertia. We therefore expect that consideration of this effect would magnify the ionization cross sections.

\begin{figure}[tbhp]
	\centering
	\includegraphics[height=.45\textwidth, angle=-90]{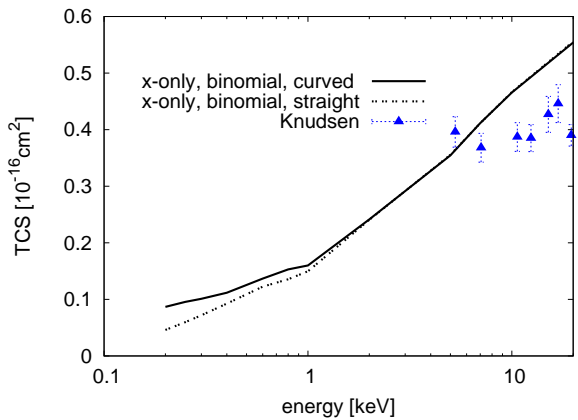}
	\caption{(Color online) Total cross section for one-fold ionization for small energies. Present results within x-only and binomial approximations and with straight-line and curved trajectories for the projectile motion. Experiment: Knudsen \cite{KnudsenTCS_1}.}
	\label{kleine_ene_1}
\end{figure}

\begin{figure}[htbp]
	\centering
	\includegraphics[height=.45\textwidth, angle=-90]{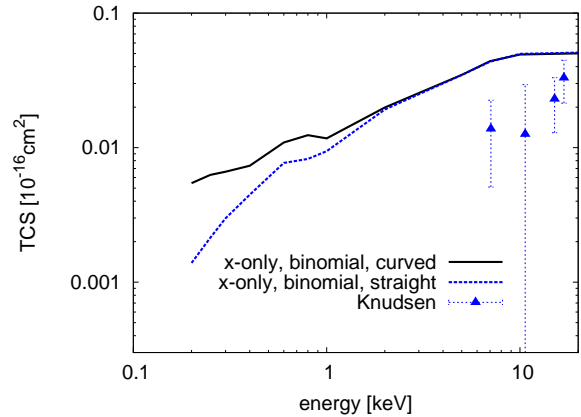}
	\caption{(Color online) Total cross section for two-fold ionization for small energies. Present results within x-only and binomial approximations and with straight-line and curved trajectories for the projectile motion. Experiment: Knudsen \cite{KnudsenTCS2}.}
	\label{kleine_ene_2}
\end{figure}

For one-fold ionization (Fig. \ref{kleine_ene_1}) as well as for two-fold-ionization (Fig. \ref{kleine_ene_2}) we find that the curved trajectory leads to a higher ionization cross section. This is due to the fact that the effective impact parameter is reduced and the effective interaction time enhanced. While the quantitative results should be taken with a grain of salt due the simple modeling of the force on the projectile, it seems clear qualitatively that the trajectory effect is significant. We reiterate that non-inertial effects have been neglected, so that the real effect of the (classical) projectile-target interaction might be even larger.

\section{Conclusions}
In this paper we have applied several approximations within TDDFT to the antiproton-helium collision system. We conclude that the correlation potential from Ref.~\cite{LK} is not suited for this system because its main characteristic, the discontinuity at $I=1$ does not come into play. Hence, until a better approximation is found, the x-only approximation should be preferred for calculations within TDDFT.
We saw that functional correlation and dynamic correlation are about equally large for one-fold ionization, whereas for two-fold ionization the functional correlation is the much more important effect. In the case of single ionization all those effects are negligible for sufficiently high energies. In order to test different approximations for the correlation potential, one should therefore focus on energies below 100\,keV. Since functional correlation is also important in that regime, it is the average ionization that should be used for benchmarking. 
For energies below 5\,keV we showed that there is a trajectory effect that significantly magnifies one-fold as well as two-fold ionization.

\end{document}